\documentclass[twocolumn,aps,prl,superscriptaddress,longbibliography]{revtex4-1}
\usepackage{graphicx}
\usepackage{amsmath}
\usepackage{bm}
\usepackage{amssymb}
\usepackage{bbold}		
\usepackage{hyperref}
\usepackage{epstopdf}
\usepackage{multirow}
\usepackage{dcolumn}

\def\bea{\begin{eqnarray}}
\def\eea{\end{eqnarray}}
\def\be{\begin{equation}}
\def\ee{\end{equation}}

\newcolumntype{d}{D{.}{.}{1}}

\def \lvec{(\kern-.26em(}

\begin{document}

\title{Control of Ultracold Photodissociation with Magnetic Fields}

\author{M. McDonald}
\altaffiliation{Present address:  Department of Physics, University of Chicago, 929 East 57th Street GCIS ESB11, Chicago, IL 60637, USA}
\affiliation{Department of Physics, Columbia University, 538 West 120th Street, New York, NY 10027-5255, USA}
\author{I. Majewska}
\affiliation{Quantum Chemistry Laboratory, Department of Chemistry, University of Warsaw, Pasteura 1, 02-093 Warsaw, Poland}
\author{C.-H. Lee}
\affiliation{Department of Physics, Columbia University, 538 West 120th Street, New York, NY 10027-5255, USA}
\author{S. S. Kondov}
\affiliation{Department of Physics, Columbia University, 538 West 120th Street, New York, NY 10027-5255, USA}
\author{B. H. McGuyer}
\altaffiliation{Present address:  Facebook, Inc., 1 Hacker Way, Menlo Park, CA 94025, USA}
\affiliation{Department of Physics, Columbia University, 538 West 120th Street, New York, NY 10027-5255, USA}
\author{R. Moszynski}
\affiliation{Quantum Chemistry Laboratory, Department of Chemistry, University of Warsaw, Pasteura 1, 02-093 Warsaw, Poland}
\author{T. Zelevinsky}
\email{tanya.zelevinsky@columbia.edu}
\affiliation{Department of Physics, Columbia University, 538 West 120th Street, New York, NY 10027-5255, USA}

\begin{abstract}     
Photodissociation of a molecule produces a spatial distribution of photofragments determined by the molecular structure and the characteristics of the dissociating light.  Performing this basic chemical reaction at ultracold temperatures allows its quantum mechanical features to dominate.  In this regime, weak applied fields can be used to control the reaction.  Here, we photodissociate ultracold diatomic strontium in magnetic fields below 10 G and observe striking changes in photofragment angular distributions.  The observations are in excellent qualitative agreement with a multichannel quantum chemistry model that includes nonadiabatic effects and predicts strong mixing of partial waves in the photofragment energy continuum.  The experiment is enabled by precise quantum-state control of the molecules.

\end{abstract}
\date{\today}
\maketitle

\newcommand{\w}{3.25in}

\newcommand{\Schematic}[1][\w]{
\begin{figure}[!htbp]
\includegraphics*[trim = 0in 0in 0in 0in, clip, width=3.375in]{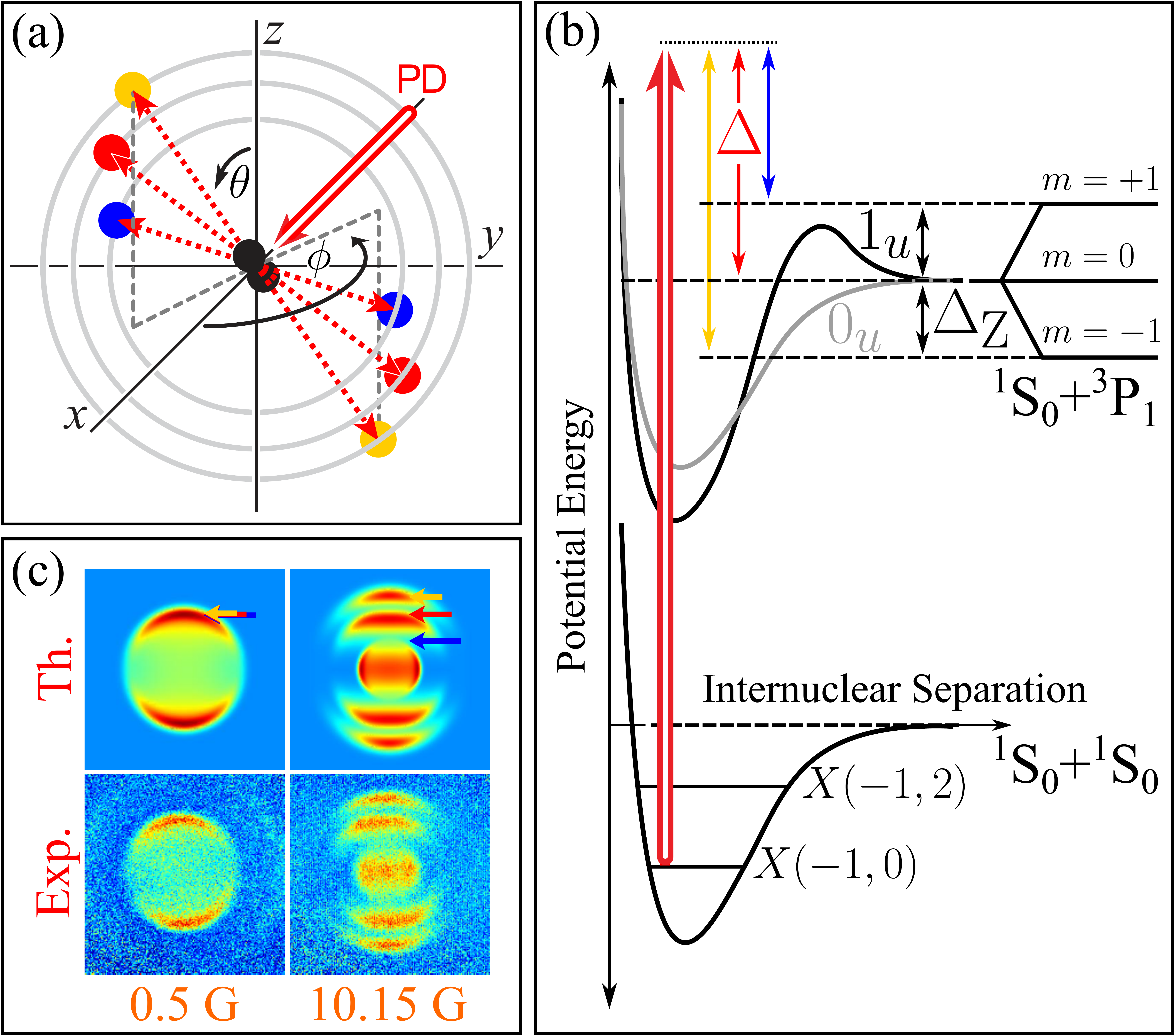}
\caption{(a)  Geometry of the photodissociation process.  The molecules are trapped in an optical lattice at the origin, while the photodissociation (PD) laser propagates along the $x$ axis.  The polar angle $\theta$ and azimuthal angle $\phi$ are defined as shown to describe the photofragment angular distributions (PADs).  The radii of the spherical shells containing the fragments after a fixed expansion time are defined by the frequency of the PD laser and the Zeeman shifts of different atomic continua.  The largest shell corresponds to a negative shift (yellow), the medium shell to an absence of shift (red), and the smallest shell to a positive shift (blue).  A camera points in the $-x$ direction and images a two-dimensional projection of the nested shells.  (b) Molecular potentials and quantum states relevant to the experiment.  The photodissociation process is designated by the double arrow.  The detuning of the PD light from the $m=0$ Zeeman component of the continuum is $\Delta$, and the symmetric Zeeman splitting has a magnitude $\Delta_Z$.  The barrier of the $1_u$ potential has a height of $\sim30$ MHz.  The numbers in parentheses are $v$ and $J_i$.  (c) An example of calculated and measured PAD images for a process where a small applied magnetic field drastically alters the outcome of the reaction.  The two pairs of images differ only by the magnitude of the applied field:  $B=0.5$ G on the left and $10.15$ G on the right.}
\label{fig:Schematic}
\end{figure}
}

\newcommand{\BfieldPD}[1][\w]{
\begin{figure}[!htbp]
\includegraphics*[trim = 0in 0in 0in 0in, clip, width=3.375in]{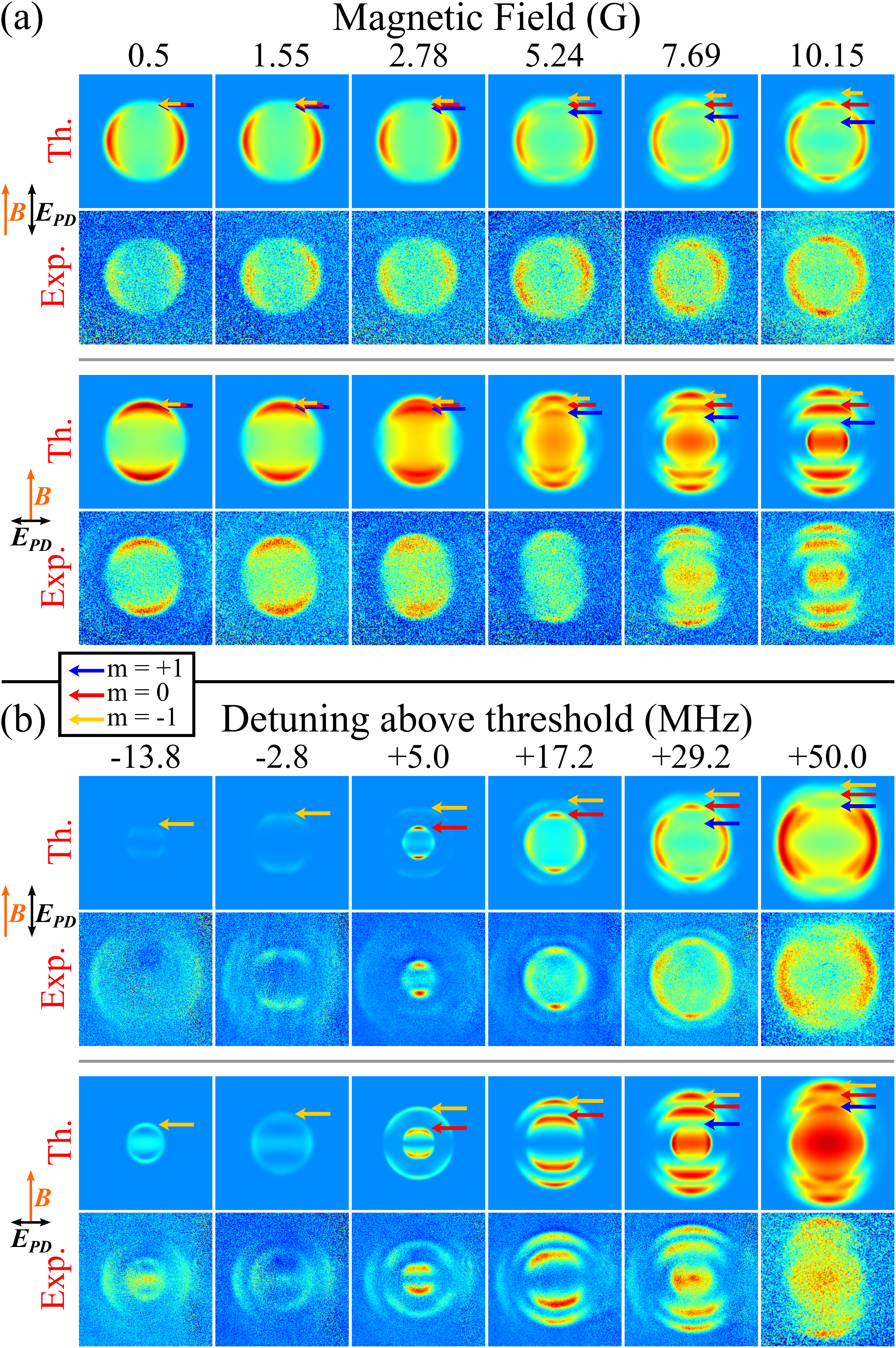}
\caption{Tuning of the photodissociation reaction with small magnetic fields, across a range of energies.  The color coding for the continuum Zeeman components is consistent between Figs. \ref{fig:Schematic} and \ref{fig:BfieldPD}.  (a) Theoretical and experimental images of PADs as the magnetic field $B$ is increased from 0.5 to 10.15 G, for the detuning $\Delta=29.2$ MHz.  (b) Theoretical and experimental PAD images at $B=10.15$ G, covering a range of $\Delta$ from -13.8 to 50.0 MHz.  As indicated in Fig. \ref{fig:Schematic}, additional channels ($m=-1$, $0$, and $1$) become available in the continuum as $\Delta$ increases, leading to extra photofragment shells.
In all experimental images, the faint outermost shell is the result of incidental photodissociation of residual $J = 2$ molecules and can be ignored.  The top and bottom pairs of rows in both (a) and (b) correspond to the light polarization parallel and perpendicular to $\vec{B}$, respectively.  Typically, 300 experiments with atoms and 300 without atoms (for background subtraction) are averaged to obtain each experimental PAD image.  The experimental images use an arbitrary brightness scale, and the relative transition strengths for different images can be inferred from this data only qualitatively.  Within each PAD, however, relative transition strengths to different $m$'s are more accurately reflected in the relative brightness of the rings.}
\label{fig:BfieldPD}
\end{figure}
}

\newcommand{\PartialWaveBMixing}[1][\w]{
\begin{figure}[!htbp]
\includegraphics*[trim = 0in 0in 0in 0in, clip, width=3.375in]{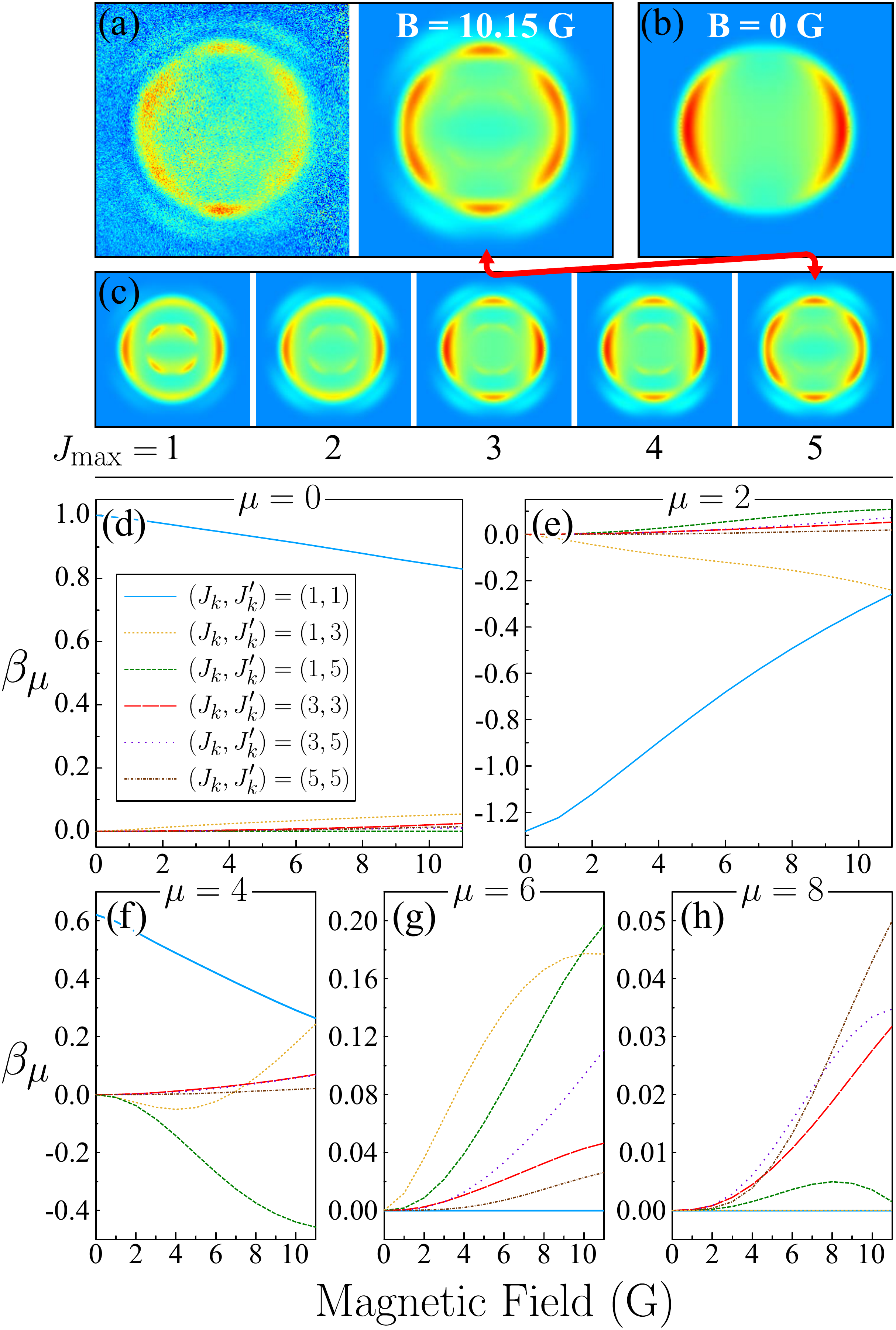}
\caption{Mixing of the continuum partial waves with a small magnetic field.  (a) Experimental and calculated PAD images for photodissociation at $B=10.15$ G.  The laser polarization is parallel to $\vec{B}$ and its detuning is $\Delta=29.2$ MHz.  The brightest fragment shell here corresponds to $m=0$.  (b) For comparison, a calculated PAD for $B=0$ consists of a simple dipolar pattern corresponding to $J=1$.  The quantum numbers $m=-1,0,1$ are unresolved.  (c) Calculated PADs of the process in (a), each image including contributions only up to the maximum partial wave $J_{\mathrm{max}}=1$ through $5$ that is included in the multichannel continuum wave function, as labeled.  Agreement with experiment is reached for $J_{\mathrm{max}}=5$.  (d-h) Plots of the anisotropy parameters $\beta_{\mu}$ versus the field strength $B$ for the dominant $m=0$ component of the PAD in (a).  The contributions of individual continuum angular-momentum pairs ($J,J'$) are shown (with $J_{\mathrm{max}}=5$), where the indexed notation corresponds to that of Eq. (10) in \cite{Note1}.}
\label{fig:PartialWaveBMixing}
\end{figure}
}

Chemical reactions at cold and ultracold temperatures exhibit quantum mechanical behavior, since at these low kinetic energies reactions possess a strong sensitivity to the details of intermolecular interactions.  Moreover, when the reactants are prepared at ultracold temperatures, their internal quantum states can be well controlled, leading to a much greater understanding of the reaction and potentially enabling a complete theoretical description.  When a reaction proceeds at such low temperatures, it becomes possible to control its outcome by applying modest electric or magnetic fields.  This occurs because the size of Stark or Zeeman shifts can be much greater than the kinetic energy \cite{BalakrishnanJCP16_UltracoldMolecules}, and the density of molecular states is high near the threshold, facilitating mixing by external fields \cite{ZelevinskyMcGuyerPRL15_Sr2ForbiddenE1}.  Field control of diatomic-molecule collisions and reactions has been investigated recently for polar molecules, largely focusing on rate constants \cite{JinNiNature10_DipolarCollisions,JinMirandaNPhys11_BimolecularReactionStereodynamics,BohnQuemenerPRA13_UltracoldMoleculeCollisionsInEBFields}.

Photodissociation of a diatomic molecule is a basic chemical reaction where a bond breaks under the influence of light.  It is related to photoassociation of an atom pair \cite{JonesRMP06} by time reversal, but has advantages for studies of ultracold chemistry.  In photodissociation, thermal averaging of the atomic collision energies is avoided and the internal and motional states of the initial molecules can be precisely engineered, leading to fully quantum-state-controlled reactions and strictly nonclassical phenomena such as matter-wave interference of the reaction products \cite{ZelevinskyMcDonaldNature16_Sr2PD}.  Here, we photodissociate ultracold diatomic strontium molecules, $^{88}$Sr$_2$, and induce dramatic changes in reaction outcomes by applying magnetic fields.  The study of photodissociation in the ultracold regime and in the presence of external fields requires us to explicitly include field-induced angular-momentum mixing into the theoretical treatment of this process.  While the theory of photodissociation has been extensively developed \cite{ZareMPC72_PhotoejectionDynamics,ShapiroBalintKurtiCP81_TriatomicPhotofragmentation,BernsteinChoiJCP86_StateSelectedPhotofragmentation,VasyutinskiiKuznetsovJCP05_PADRoleMoleculeAxisRotation} including the effects of magnetic fields \cite{BeswickCP79_PhotopredissAngDistrInBField}, previously the total angular momentum was considered a conserved quantum number.  In the regime explored here, this is no longer the case.  Combined with a multichannel quantum-chemistry molecular model \cite{MoszynskiSkomorowskiJCP12_Sr2Dynamics,KillianBorkowskiPRA14_SrPAMassScaling}, the theoretical treatment we have developed here faithfully reproduces all our experimental observations.

\Schematic
In the experiment we directly observe and record the photofragment angular distributions (PADs) in the millikelvin energy regime.  The molecules are prepared at microkelvin temperatures in an optical lattice, and are subsequently fragmented with laser light \cite{ZelevinskyMcDonaldNature16_Sr2PD}.  The one-dimensional lattice is a standing wave of far-off-resonant light at 910 nm and is approximately 1 MHz (or 50 $\mu$K) deep.  The geometry of the setup is defined in Fig. \ref{fig:Schematic}(a).  Photodissociation results in two counter-propagating photofragments, an atom in the ground state $^1S_0$ and an atom in the electronically excited state $^3P_1$ which decays to $^1S_0$ with a 10 $\mu$s lifetime.  These atoms are absorption imaged using a charge-coupled device camera on the strong Sr transition at 461 nm.  The imaging light is turned on for a short duration of $\sim10$ $\mu$s, at a time $\tau$ (between 250 and 600 $\mu$s) after the 20-50 $\mu$s photodissociation pulse at 689 nm.  During this time, the photofragments freely expand and effectively form spherical shells with radii determined by the frequency of the photodissociation light and the Zeeman shifts of the atomic continua.  The camera is nearly on-axis with the lattice, thus capturing a two-dimensional projection of the spherical shells since the atoms effectively originate from a point source.  The laboratory quantum axis points along the applied magnetic field $\vec{B}$, which has a vertical orientation that defines the polar angle $\theta$ and azimuthal angle $\phi$.  The dependence of the photofragment density on these angles is our key observable and encodes the quantum mechanics of the reaction.  The photodissociation light polarization is set to be either vertical or horizontal.

Figure \ref{fig:Schematic}(b) illustrates the Sr$_2$ molecular structure relevant to this work.  The molecules are created from ultracold atoms via photoassociation \cite{ZelevinskyReinaudiPRL12_Sr2} in the least-bound vibrational level, denoted by $v=-1$, of the electronic ground state X$^1\Sigma_g^+$ (correlating to the $^1S_0{+^1S_0}$ atomic threshold).  Initially, the molecules occupy two rotational states with the total angular-momentum quantum numbers $J_i=0$ and $2$, but the $J_i=2$ population is mostly removed prior to fragmentation by a laser pulse resonant with an excited molecular state.  The $J_i=0$ molecules (with a projection quantum number $M_i=0$) are coupled by the photodissociation laser to the singly-excited continuum above the $0_u$ and $1_u$ \textit{ungerade} potentials (correlating to the $^1S_0{+^3P_1}$ atomic threshold), where the numbers refer to the total atomic angular momentum projections onto the molecular axis.  Under an applied field $B>0$, the atomic energy levels split by the Zeeman interaction into the $m=-1$, $0$, and $1$ sublevels, where the energy separation between the neighboring sublevels is $h\Delta_Z=1.5\mu_BB$ and $\mu_B$ is the Bohr magneton.  The radius of each photofragment shell is $v\tau$ where the velocity $v=\sqrt{h(\Delta-m\Delta_Z)/m_{\mathrm{Sr}}}$, $h$ is the Planck constant, $\Delta$ is the frequency detuning of the photodissociation light from the $m=0$ component of the continuum, and $m_{\mathrm{Sr}}$ is the atomic mass of Sr.

If the photodissociation laser detuning is large and negative, $\Delta<-\Delta_Z$, no photofragments should be detectable because the target energy is below the lowest threshold.  If the detuning is small and negative, $-\Delta_Z<\Delta<0$, then only one fragment shell should be visible, corresponding to $m=-1$.  If the detuning is small and positive, $0<\Delta<\Delta_Z$, we expect to observe two fragment shells, with $m=-1$ and $0$.  Finally, if the detuning is large and positive, $\Delta_Z<\Delta$, we expect three fragment shells with all possible values of $m$.  This is the case in the example of Fig. \ref{fig:Schematic}(c) that shows a strong alteration of the PAD for $B=10.15$ G compared to 0.5 G.
Here we make the distinction between the angular-momentum projection quantum numbers $m$ and $M$, the latter denoting the projection of the total angular momentum $J$ in the continuum.  Electric-dipole (E1) selection rules require $M=M_i=0$ if the photodissociation laser polarization is parallel to the quantum axis and $M=\pm1$ if the polarization is perpendicular.  In contrast, there are no such selection rules for the atomic magnetic sublevels $m$ which can be superpositions of several $M$.

When a $J_i=0$ diatomic molecule is photodissociated via a one-photon E1 process without an applied field, we expect and observe a dipolar-shaped PAD with an axis set by the laser polarization \cite{ZelevinskyMcDonaldNature16_Sr2PD}, as in the nearly field-free case of Fig. \ref{fig:Schematic}(c).  This can be understood either by visualizing a spherically symmetric molecule absorbing light with a dipolar probability distribution, or by applying angular-momentum selection rules that require $J=1$ for the outgoing channel, which has a dipolar angular distribution with a single spatial node.  We find that with a nonzero $B$ this is no longer the case, and instead observe complicated structures with multiple nodes.

\BfieldPD
The main results of the experiment and theory are summarized in Fig. \ref{fig:BfieldPD}.
The two-dimensional projections of the PADs onto the imaging plane, with the detuning $\Delta=29.2$ MHz, are shown in Fig. \ref{fig:BfieldPD}(a) for a progression of magnetic fields $B$ from 0.5 to 10.15 G.  The removal of the $J_i=2$ molecules is imperfect which results in the faint outermost shell that can be ignored.  The top pair of rows corresponds to parallel light polarization and the bottom pair to perpendicular polarization.  We observe a transformation from simple dipolar patterns at $B=0$ to more complex patterns that exhibit a multiple-node structure at 10.15 G.  Figure \ref{fig:BfieldPD}(b) shows PADs that are observed when $B$ is kept fixed at 10.15 G while $\Delta$ is varied from -13.8 to 50.0 MHz, again for both cases of linear light polarization.
For the entire range of continuum energies, we observe PADs that exhibit a multinode structure.  As $\Delta$ and $B$ are varied, the angular dependence, or anisotropy, of the outgoing PAD is strongly affected.  The zero-field evolution of the PADs with energy for this continuum is discussed in detail in \cite{ZelevinskyMcDonaldNature16_Sr2PD}.
All additional features observed here are due to the continuum partial waves $J$ being strongly mixed by the applied field.

As Fig. \ref{fig:BfieldPD} demonstrates, our theoretical results are in excellent qualitative agreement with the experimental data.  The theory involves extending the standard treatment of diatomic photodissociation to the case of mixed angular momenta in the presence of a magnetic field, and applying it to the quantum-chemistry model of the $^{88}$Sr$_2$ molecule \cite{MoszynskiSkomorowskiJCP12_Sr2Dynamics,KillianBorkowskiPRA14_SrPAMassScaling}.  As detailed in \footnote{See Supplemental Material for the quantum mechanical description of photodissociation in applied magnetic fields.}, the PADs can be described by the expansion
\begin{align}
I(\theta, \phi) \propto   \beta_0\left(1 + \sum_{\mu=1}^{\infty} \sum_{\nu = 0}^{\mu} \beta_{\mu \nu} P_{\mu}^{\nu} (\cos \theta) \cos (\nu \phi)\right)
\label{Eq:PADparametrization}
\end{align}
where $P_{\mu}^{\nu} (\cos \theta)$ are the associated Legendre polynomials and the $\beta_{\mu \nu}$ coefficients are called anisotropy parameters.  In the case of parallel light polarization, the PADs are cylindrically symmetric (no $\phi$ dependence) \cite{ZelevinskyMcDonaldNature16_Sr2PD}, and we set $\beta_{\mu}\equiv\beta_{\mu0}$ while all other $\beta_{\mu\nu}$ vanish.  The $\mu$ are even for homonuclear dimers.  The anisotropy parameters in expression (\ref{Eq:PADparametrization}) can be evaluated from Fermi's golden rule after properly representing the initial (bound-state) and final (continuum) wave functions, including mixing of the angular momenta $J_i$ and $J$ by the magnetic field.

\PartialWaveBMixing
The most salient feature of ultracold photodissociation in nonzero magnetic fields is the dramatic change of the PADs which tend to become significantly more complex as the field is increased.  Figure \ref{fig:PartialWaveBMixing}(a,b) compares the photodissociation outcome at $B=10.15$ G (also in the top right of Fig. \ref{fig:BfieldPD}(a)) to that at $B=0$.  Besides the appearance of an outer shell caused by Zeeman splitting in the continuum, the central $m=0$ shell gains additional lobes as compared to the purely dipolar ($J=1$) pattern for $B=0$.  This effect arises from the magnetic field admixing higher partial waves in the continuum, as the density of states is particularly high near the dissociation threshold \cite{ZelevinskyMcGuyerPRL15_Sr2ForbiddenE1}.  We show this directly by simulating the image of the PAD on the right panel of Fig. \ref{fig:PartialWaveBMixing}(a) while using a series of cutoff partial waves $J_{\mathrm{max}}$ that are included in the continuum wave function.  The result is in Fig. \ref{fig:PartialWaveBMixing}(c).  The PAD evolves with increasing $J_{\mathrm{max}}$, only reproducing the data at $J_{\mathrm{max}}=5$.  We have confirmed that increasing $J_{\mathrm{max}}$ further does not alter the PAD appreciably.
($_{\mathrm{max}}=1$ if $B=0$.)  The evolution of the PADs with increasing magnetic field can be alternatively described by plotting the anisotropy parameters $\beta_{\mu}$ as functions of $B$.  Figure \ref{fig:PartialWaveBMixing}(d-h) shows this for the PAD in Fig. \ref{fig:PartialWaveBMixing}(a), for anisotropies of order $\mu=0$ through $8$ that we can resolve in the experiment.  The curves correspond to contributions from pure and mixed exit-channel partial waves of Eq. (10) in \cite{Note1}, with $J_k,J_k'$ varying from 1 to 5.  These plots directly show that higher-order anisotropy ($\mu>4$) arises already at $\sim1$ G and is dominated by the admixing of increasingly higher angular momenta in the continuum.  Note in the plots of Fig. \ref{fig:PartialWaveBMixing}(d-h) that if $B=0$, the maximum anisotropy order is $\mu=4$ for our quantum numbers.

We have shown that the chemical reaction of photodissociation can be strongly altered in the ultracold regime by applied magnetic fields.  In this work, the fragmentation of $^{88}$Sr$_2$ molecules was explored for a range of fields from 0 to 10 G, and for a variety of energies above threshold in the 0--2 mK range.  The near-threshold continuum has a high density of partial waves that are readily mixed by the field, resulting in pronounced changes of the photofragment angular distributions.  The theory of photodissociation, after explicit accounting for field-induced mixing of angular momenta in the bound and continuum states, and combined with an accurate quantum-chemistry molecular model, has yielded excellent agreement with experimental data.  The experiment and its clear interpretation was made possible by preparing the molecules in well-defined quantum states.  We have shown that ultracold molecule techniques allow a high level of control over basic chemical reactions with weak applied fields.  Moreover, this work serves as a test of \textit{ab initio} molecular theory in the continuum.

We acknowledge the ONR Grants No. N00014-16-1-2224 and N00014-17-1-2246, the NSF Grant No. PHY-1349725, and the NIST Grant No. 60NANB13D163 for partial support of this work.  R. M. and I. M. also acknowledge the Polish National Science Center Grant No. 2016/20/W/ST4/00314 and M. M. the NSF IGERT Grant No. DGE-1069240.

\section{Supplemental Material}

This supplement summarizes our extension of the quantum mechanical theory of photodissociation to the situation where the total angular momentum is not a conserved quantum number, as is the case in our ultracold-molecule experiments with applied magnetic fields.

\subsection{Notation}

In our theoretical description of quantum mechanical photodissociation, the notation follows \cite{VasyutinskiiKuznetsovJCP05_PADRoleMoleculeAxisRotation,VasyutinskiiShterninJCP08_PhotofragmentsCoriolis} with slight changes.  The main symbols are as follows:
\begin{itemize}
\itemsep-0.4em
\item $\mathbf{R}=(R,\Theta,\Phi)$:  vector that connects the pair of atomic fragments.  The angles $\Theta,\Phi$ are defined relative to the molecular axis.
\item $\{\mathbf{r}\}$:  set of electronic coordinates of the atoms.
\item $\mathbf{k}=(k,\theta,\phi)$:  scattering wave vector of the photofragments.  The angles $\theta,\phi$ are defined relative to the $z$ axis, the quantum axis in the lab frame.
\item $\mathbf{j}=\mathbf{j}_1+\mathbf{j}_2$:  combined angular momentum of the atomic fragments.
\item $m_j\equiv m$:  projection of $\mathbf{j}$ onto the lab $z$ axis.
\item $\mathbf{l}$:  orbital angular momentum of the atomic fragments about their center of mass.
\item $m_l$:  projection of $\mathbf{l}$ onto the lab $z$ axis.
\item $\mathbf{J}=\mathbf{j}+\mathbf{l}$:  total angular momentum of the photodissociated system.
\item $M=m_j+m_l$:  projection of $\mathbf{J}$ onto the lab $z$ axis.
\item $\Omega$:  projection of $\mathbf{J}$ onto the molecular axis.
\end{itemize}

A novel aspect of this work is that the total angular momentum $J$ is not conserved, while $M$ is the rigorously conserved quantum number.  To account for this, we introduce the indexed angular momenta $J_R$ and $J_k$, where the subscripts $R$ and $k$ denote the entrance and exit channels for the continuum wave function of the photofragments.

\subsection{Parametrization of the photofragment angular distribution}

For photodissociation of a diatomic molecule, the photofragment angular distribution (PAD) is given by the intensity function of the polar angle $\theta$ and azimuthal angle $\phi$ as
\begin{align}
 I(\theta, \phi) \propto   \beta_0 \left(1+ \sum_{\mu=1}^{\infty} \sum_{\nu = 0}^{\mu} \beta_{\mu \nu} P_{\mu}^{\nu} (\cos \theta) \cos (\nu \phi)\right),
\tag{\ref{Eq:PADparametrization} revisited}
\end{align}
where $P_{\mu}^{\nu} (\cos \theta)$ are the associated Legendre polynomials and $\beta_{\mu \nu}$ are anisotropy parameters.
For the parallel polarization of light, the PADs are cylindrically symmetric (no $\phi$ dependence) \cite{ZelevinskyMcDonaldNature16_Sr2PD}, and we set $\beta_{\mu}\equiv\beta_{\mu0}$ while all other $\beta_{\mu\nu}$ vanish.

\subsection{Theory of photodissociation in a magnetic field}

The photodissociation process is characterized by a differential cross section $\sigma(\hat{k})=|f|^2$, defined by Fermi's golden rule with the electric-dipole (E1) transition operator. The corresponding scattering amplitude is
\begin{align}
 \label{fermi}
f\propto\langle \Psi_\textbf{k}^{p_f jm_j}(\{\textbf{r}\},\textbf{R}) |  \hat{T}^1_{\text{E1}}  | \Psi ^{p_i}_{J_iM_i} (\{\textbf{r}\}, \textbf{R} )  \rangle,
\end{align}
where $\Psi ^{p_i}_{J_iM_i} (\{\textbf{r}\}, \textbf{R} ) $ and $\Psi_\textbf{k}^{p_f jm_j}(\{\textbf{r}\},\textbf{R})$ are the initial (bound-staet) and final (continuum) wave functions.
This description was first applied in the Born-Oppenheimer approximation in \cite{ZareMPC72_PhotoejectionDynamics}.  Furthermore, the treatment of triatomic photodissociation \cite{ShapiroBalintKurtiCP81_TriatomicPhotofragmentation,ShapiroBalintKurti_QuantTheoryPhotodissociation} is useful for our diatomic case with additional internal atomic structure.  Detailed derivation of photodissociation theory for individual magnetic sublevels is available in literature \cite{PowisUnderwoodJCP00_PDPolarizedDiatomics,VasyutinskiiKuznetsovJCP05_PADRoleMoleculeAxisRotation,VasyutinskiiShterninJCP08_PhotofragmentsCoriolis}.  However, to the best of our knowledge, the wave functions in the presence of a magnetic field (eigenfunctions of the Zeeman Hamiltonian) have not been previously incorporated into the theory.  Photodissociation in a magnetic field was discussed in \cite{BeswickCP79_PhotopredissAngDistrInBField}, but $J$ was assumed to be a good quantum number, which is not the case in our experimental regime even for weak fields.

In this work we consider E1 photodissociation of weakly bound ground-state $^{88}$Sr$_2$ molecules into the $0_u^+/1_u$ \textit{ungerade} continuum correlating to the $^1S_0{+^3P_1}$ atomic threshold.  The transition operator connecting the initial and final wave functions is assumed to be constant and proportional to the atomic value.  This approximation is valid for weakly bound molecules.
It is assumed that the field affects only the excited states, since the ground state (correlating to $^1S_0{+^1S_0}$) is nearly nonmagnetic.

\subsubsection{Bound state wave function}

Since the initial (bound-state) wave function is not affected by the magnetic field $B$, it is given by the standard form using the electronic coordinates $\{\textbf{r}\}$ and the internuclear vector $\textbf{R}=(R,\theta,\phi)$,
\begin{widetext}
\begin{align}
\label{Psi-initial}
\Psi ^{p_i}_{J_iM_i} (\{\textbf{r}\},\textbf{R})   &= \frac{1}{\sqrt{2}} \sqrt{\frac{2J_i+1}{8 \pi^2}} \sum_{\Omega_i=-J_i}^{+J_i} \frac{1}{\sqrt{1+\delta_{\Omega_i 0}}} \\ & \nonumber
 \times \left( D^{(J_i) ^\star}_{M_i \Omega_i} (\phi,\theta,0) \psi^{p_i}_{J_i \Omega _i} (\{\textbf{r}\},R) + \sigma_i D^{(J_i) ^\star}_{M_i -\Omega_i} (\phi,\theta,0) \psi^{p_i}_{J_i -\Omega _i} (\{\textbf{r}\},R) \right) ,
\end{align}
\end{widetext}
where the subscript $i$ denotes the initial molecular state, $\sigma$ is the spectroscopic parity defined as $p(-1)^J$, $p$ is the parity with respect to the space-fixed inversion, and $D^J_{M\Omega}$ are the Wigner rotation matrices.
In Hund's case (c) the internal wave function $\psi^{p_i}_{J_i \Omega _i} (\{\textbf{r}\},R)$ can be
represented by the Born-Huang expansion \cite{BornHuang,MoszynskiBusseryHonvaultJCP06_Ca2AbInitio},
\begin{equation}
\psi^{p_i}_{J_i \Omega _i} (\{\textbf{r}\},R)=\sum_{n_i}\phi_{n_i\Omega_i}(\{\textbf{r}\};R)\chi_{n_iJ_i \Omega _i}^{p_i}(R),
\label{psi-internal}
\end{equation}
where $\phi_{n_i\Omega_i}(\{\textbf{r}\};R)$ are the solutions of the electronic
Schr\"odinger equation including spin-orbit coupling,
$\chi_{n_iJ_i \Omega _i}^{p_i}(R)$
are the rovibrational wave functions, and the index $n$ labels all relativistic electronic channels that are included in the model.
The rovibrational wave functions are solutions of a system of coupled differential equations as detailed in \cite{MoszynskiSkomorowskiJCP12_Sr2Dynamics}.

\subsubsection{Continuum state wave function}

The correct description of the final (continuum) wave function is crucial to explaining and predicting the outcome of photodissociation in a magnetic field.  Zeeman mixing of rovibrational levels 
was responsible for observations of forbidden molecular (bound-to-bound) E1 transitions that violate the $\Delta J = 0, \pm 1$ selection rule \cite{ZelevinskyMcGuyerPRL15_Sr2ForbiddenE1}.  Similar effects are expected for bound-to-continuum photodissociation transitions.  

In a magnetic field, the only conserved quantities are the projection of the total angular momentum $M$ and the total parity.
For dissociation to the \textit{ungerade} continuum ($^1S_0{+^3P_1}$), the atomic angular-momentum quantum number is $j=1$.  Its magnetic sublevels $m_j = 1, 0, -1$ are split by the field, and therefore the photodissociation cross-section calculations have to be performed for each sublevel individually.
The $0_u^+$ and $1_u$ states, corresponding to $\Omega =0$ and $|\Omega| = 1$, are coupled by the nonadiabatic Coriolis interaction, while the Zeeman interaction couples the $\Delta J = 0, \pm 1$ states.  For these reasons, two sets of additional numbers are introduced:  $\Omega_R$, $J_R$, $n_R$ correspond to the entrance channels of the multichannel continuum wave function and $\Omega_k$, $J_k$, $n_k$ correspond to the exit channels.  In this work, selection rules fix $\Omega_R=1$.

As a result, the continuum wave function corresponding to the wave vector $\textbf{k}$ is
\begin{widetext}
\begin{align}
\label{Psi-final}
 \Psi_\textbf{k}^{p jm_j}(\{\textbf{r}\},\textbf{R})
  = &
 \sum_{ J_kM J_R}
  \sum_{\Omega_k \Omega_R}  \sum_{l m_l} (-1)^{J_k + \Omega_k} Y ^{\star}_{l m_l}(\hat{k}) \sqrt{\frac{2l+1}{8\pi^2}}
  \begin{pmatrix}
  J_k & j & l \\
  -M &m_j& m_l
 \end{pmatrix}
  \begin{pmatrix}
  J_k & j & l\\
  \Omega_k & -\Omega_k& 0
 \end{pmatrix}
 \\ \nonumber
 &
 \times
 \frac{\sqrt{2J_k+1} }{\sqrt{1+\delta_{\Omega_k 0}}}
 \frac{1}{\sqrt{1+\delta_{\Omega_R 0}}}
\left(D^{J_R}_{\Omega_RM}(\hat{R})  \psi^{jJ_k\Omega_k p}_{J_R\Omega_R} (\{\textbf{r}\},R) + p  D^{J_R}_{-\Omega_RM}(\hat{R}) \psi^{jJ_k\Omega_k p}_{J_R - \Omega_R} (\{\textbf{r}\},R)\right),
\end{align}
\end{widetext}
where $Y_{lm_l}$ are the spherical harmonics.
The detailed derivation of this wave function is found in \cite{ShapiroBalintKurtiCP81_TriatomicPhotofragmentation,VasyutinskiiShterninJCP08_PhotofragmentsCoriolis}, but for the simpler case when $J$ is a good quantum number and the $J_k, J_R$ channel numbers are not needed.

The function  $\psi^{jJ_k\Omega_k p}_{J_R\Omega_R} (\{\textbf{r}\},R) $ can be expressed by the Born-Huang expansion as
\begin{align}
\label{born-huang}
  \psi^{jJ_k\Omega_k p}_{J_R\Omega_R} (\{\textbf{r}\},R) = \sum_{n_k n_R} \phi_{n_R \Omega_R}(\{\textbf{r}\},R)\chi^{jJ_k\Omega_k n_k p}_{J_R\Omega_R n_R} (R ),
\end{align}
where the wave functions $\phi_{n_R\Omega_R}(\{\textbf{r}\},R)$ are the solutions of the electronic Schr\"{o}dinger equation.
The rovibrational wave functions $\chi^{jJ_k\Omega_k n_k p}_{J_R\Omega_R n_R} (R )$ are obtained by solving the nuclear Schr\"{o}dinger equation with the Hamiltonian
\begin{align}
\label{hamiltonian}
  \hat{H} = -\frac{\hbar^2}{2 \mu R^2} \frac{\partial}{\partial R}R^2\frac{\partial}{\partial R} + \frac{\hbar^2\textbf{l}^2}{2 \mu R^2} + V(R),
\end{align}
where $V(R)$ is the potential matrix including the Coriolis and Zeeman couplings, $\hbar$ is the reduced Planck constant, and $\mu$ is the reduced atomic mass.

At large interatomic distances in the presence of external fields, the asymptotic value of $V(R)$, or $V_{{\rm as}}$, is not diagonal in the basis of the wave functions $\phi_{n_R\Omega_R}(\{\textbf{r}\},R)$.  It is then necessary to introduce a transformation $\mathbb{C}$ that diagonalises $V_{{\rm as}}$.  The rovibrational functions $\chi^{jJ_k\Omega_k n_k p}_{J_R\Omega_R n_R} (R )$ form a matrix $\bar{\mathbb{X}}$ that is propagated to large distances and transformed to the basis that diagonalizes the asymptotic potential, $\mathbb{X} = \mathbb{C}^T \bar{\mathbb{X}} \mathbb{C}$.  Then the boundary conditions are imposed \cite{Levine,JohnsonJComputPhys73_MultichannelLogDerivMethod} as
\begin{align}
\label{boundary-conditions}
 \mathbb{X}(R) \rightarrow J(R) + N(R) \cdot K,
\end{align}
where $K$ is the reaction matrix, $J(R)$ and $N(R)$ are the diagonal matrices containing the spherical Bessel functions for the open channels,
\begin{align}
 [J(R)]_{ij} = \delta_{ij} \frac{1}{\sqrt{k_j }} j_l(k_j R), \quad
 [N(R)]_{ij} = \delta_{ij} \frac{1}{\sqrt{k_j }} n_l(k_j R),
\end{align}
$k_j$ is the wave number of the $j$th channel, and $l$ is the orbital angular momentum of the $j$th channel.
A more detailed description of the close-coupled equations in a magnetic field can be found in \cite{DalgarnoKremsJCP04_MoleculeCollisionsBFieldSpinDepolariz}.

\subsubsection{Anisotropy parameters}

After inserting the wave functions (\ref{Psi-initial}) and (\ref{Psi-final}) into Fermi's golden rule (\ref{fermi}) and transforming the cross section for the photodissociation process using the Clebsch-Gordan series and properties of the Wigner $3j$ symbols, we get the following expansion for the PAD:
\begin{align}
 I(\theta, \phi) \propto   \beta_0 \left(1+ \sum_{\mu=1}^{\infty} \sum_{\nu = 0}^{\mu} \beta_{\mu \nu} P_{\mu}^{\nu} (\cos \theta) \cos (\nu \phi)\right),
 \tag{\ref{Eq:PADparametrization} revisited}
\end{align}
where the anisotropy parameters are given by
\begin{widetext}
\begin{align}
\label{beta}
 \beta_{\mu \nu} = &  \frac{1}{\beta_0}\sum_{J_k J_R  J'_k J'_R  } \sum_{l  l' M M'} t^{J_k}_{J_R} t^{J'_k\star}_{J'_R}  U^{J_k l }_{ J_R M} U^{J_k' l' }_{ J_R'M'} [\mu] \sqrt{\frac{(\mu - \nu)!}{(\mu + \nu )!}} (2-\delta_{M M'})
 \begin{pmatrix}
  l & l' & \mu\\
  M - m_j & m_j - M' & \nu
 \end{pmatrix}
  \begin{pmatrix}
  l & l' & \mu \\
 0 & 0 & 0
 \end{pmatrix}
\end{align}
\end{widetext}
and $[A]\equiv2A+1$.
The symbols $U^{J_k l }_{ J_R M}$ in Eq. (\ref{beta}) are defined as
\begin{widetext}
\begin{align}
\label{u-coef}
 U^{J_k l }_{ J_R M} = & \sum_{P \Omega_k m_l }  (-1)^{J_k + \Omega_k - m_j}
  [l]
 \frac{\sqrt{[J_k]} }{\sqrt{1+\delta_{\Omega_k 0}}}
 \begin{pmatrix}
  J_k & j & l \\
  -M &m_j& m_l
 \end{pmatrix}
 \begin{pmatrix}
  J_k & j & l\\
  \Omega_k & -\Omega_k& 0
  \end{pmatrix}
  \begin{pmatrix}
  J_R & 1 & J_i\\
  -M &P& M_i
 \end{pmatrix},
\end{align}
\end{widetext}
and the symbols $t^{J_k}_{J_R}$ are the scaled matrix elements of the asymptotic body-fixed E1 transition operator $d_{\mathrm{BF}}$ with the initial and final rovibrational wave functions,
\begin{widetext}
\begin{align}
\label{t-coef}
 t^{J_k }_{J_R} = & \frac{1}{2 \sqrt{2}}
  \sum_{\Omega_i=-J_i}^{J_i} \sum_{\Omega_R =-J}^{J} \sum _{q=-1}^{1} \sum_{n_k n_R} (-1)^{M-\Omega_R} \frac{\sqrt{[J_i]}}{\sqrt{1+\delta_{\Omega_i 0}}}  \frac{\sqrt{[J_R]}}{ \sqrt{ 1+\delta_{\Omega_R 0}}}   \\ \nonumber
\times  &
\begin{pmatrix}
  J_R & 1 & J_i\\
  -\Omega_R &q& \Omega_i
 \end{pmatrix}
\langle  \chi^{jJ_k\Omega_k n_k p}_{J_R\Omega_R n_R} (R )  | d_{\mathrm{BF}} | \chi_{n_iJ_i \Omega _i}^{p_i}(R)  \rangle.
\end{align}
\end{widetext}
The normalization factor $\beta_0$ is given by
\begin{align}
\label{beta0}
 \beta_0 = \sum_{l M}\left|\sum_{J_k J_R}t^{J_k }_{J_R} U^{J_k l }_{ J_R M}\sqrt{2l+1}(-1)^{J_k}\right|^2.
\end{align}
In Eq. (\ref{u-coef}),
the polarization index $P=0$ if the photodissociation light is polarized along the $z$ axis, while $P=-1,1$ if the light is polarized perpendicularly to the $z$ axis.

The properties of the $3j$ symbols force the following rules for the $\mu, \nu$ indices:
\begin{itemize}
\itemsep-0.4em
\item $\mu$ is even for homonuclear dimers.
\item $\mu_{\mathrm{max}} =2J_{k,\mathrm{max}}+2j= 2 J_{k,\mathrm{max}} +2$ for resolved $m$ sublevels. Thus the number of terms in the expansion (\ref{Eq:PADparametrization}) is limited by the number of channels used to construct the continuum wave function.  (When the $m$ sublevels are degenerate and are observed simultaneously, additional symmetry leads to $\mu_{\mathrm{max}} =2J_{k,\mathrm{max}}$.)  If $B=0$, then $J_{k,\mathrm{max}}=1$ and $\mu_{\mathrm{max}}=4$, as can be seen in Fig. 3(d-h) of the manuscript.
\item $\nu=M'-M$.  Since $M=M'=M_i$ for parallel light polarization, $\nu =0$ and thus the photodissociation cross section is cylindrically symmetric.
\end{itemize}

The anisotropy parameters presented in Fig. 3(d-h) of the manuscript are calculated using Eq. (\ref{beta}), but instead of summing over $J_k, J_k'$, the contributions of each combination $J_k, J_k'$ are individually plotted, and subsequently divided by $\beta_0$ from Eq. (\ref{beta0}).

%

\end{document}